\documentclass[a4paper,twoside]{ilcws08}
\usepackage[latin1]{inputenc}
\usepackage[dvips]{graphicx}
\usepackage{wrapfig,rotating}
\usepackage{array,graphicx,tabularx,color}
\usepackage{amscd,amssymb}

\def\zh2nnh{$ZH \to \nu\bar{\nu}H$}
\def\zh2qqh{$ZH \to q\bar{q}H$}
\def\zh2qqcc{$ZH \to q\bar{q}c\bar{c}$}

\title{\bf Study of the Higgs Direct Reconstruction in $ZH \to q\bar{q}H$ for ILC}

\author{
 \textbf{Hiroaki Ono\thanks{TEL:+81-25-267-1500-(537), MAIL:ono@ngt.ndu.ac.jp}}\\\\ 
 \textit{Nippon Dental University School of Life Dentistry at Niigata, Niigata, Japan}
} 

\date{\today}

\begin{document}

\maketitle

\begin{abstract}
 Precise measurement of the Higgs boson properties is an important issue of the
 International Linear Collider (ILC) experiment.
 We studied the accuracy of the Higgs mass reconstruction in the $ZH \to q\bar{q}H$ multi-jet process
 with the Higgs mass of $M_{H}=120~{\rm GeV}$ at $\sqrt{s}=250~{\rm GeV}$ with the ILD detector model.
 In this study,
 we obtained the reconstructed Higgs mass of $M_{H} = 120.79 \pm 0.089~{\rm GeV}$
 and 5.3\%  measurement accuracy of the cross-section for $ZH \to q\bar{q}b\bar{b}$
 with the integrated luminosity of $\mathcal{L}=250~{\rm fb^{-1}}$ data samples.
 \end{abstract}


 \section{Introduction}

\begin{wrapfigure}{r}{0.4\textwidth}
  \begin{center}
  \includegraphics[width=0.38\textwidth]{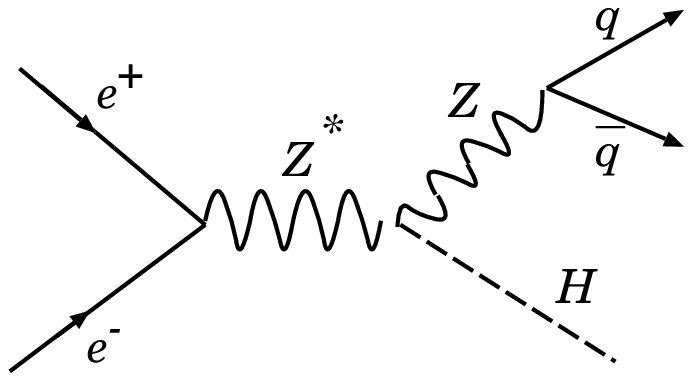}
  \caption{Higgs boson production via Higgs-strahlung $(ZH)$ process and $Z$ mainly decay to $q\bar{q}.$}
  \label{fig:ZH_diagram}
  \end{center}
 \end{wrapfigure}

\begin{figure}[htbp]
  \begin{center}
   \includegraphics[width=\textwidth]{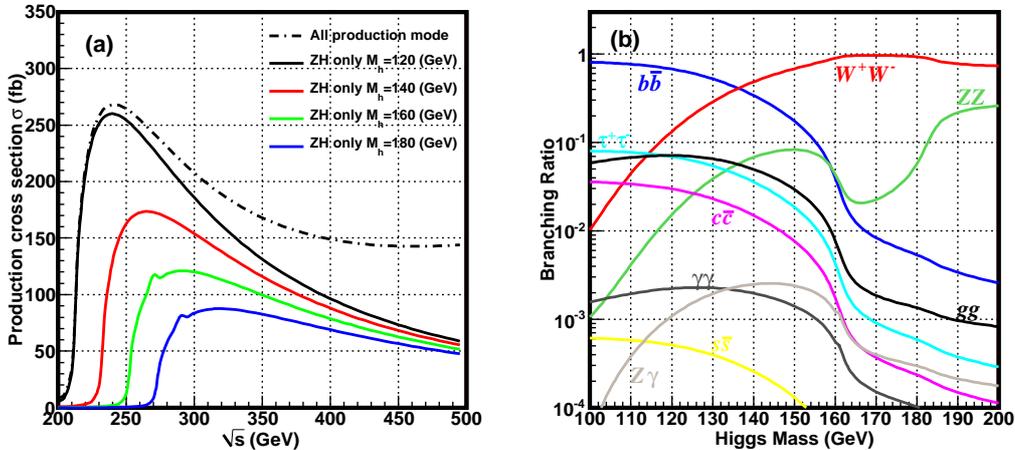}
   \caption{(a). Production cross-section of the Higgs boson as a function of center-of-mass energy ($\sqrt{s}$)
   and (b). branching ratio of the Higgs decay as a function of the Higgs mass.}
   \label{fig:ZH_xsec}
  \end{center}
 \end{figure}

 International Linear Collider (ILC)~\cite{RDR} is a future $e^{+}e^{-}$ collider experiment
 for the precise measurement and the validation of
 the Standard Model (SM) physics,
 especially for the measurement of the Higgs boson property,
 even the discovery of the Higgs boson will be realized in Large Hadron Collider (LHC) experiment.
 In the SM, light Higgs boson mass ($M_{H}$) is predicted around the
 $114.4~{\rm GeV} \leq M_{H} \leq 160~{\rm GeV}$ from the study
 in LEP~\cite{LEP} and Tevatron~\cite{CDF} experiment.
 The largest production cross-section for SM Higgs boson is obtained
 through the Higgs-strahlung ($e^{+}e^{-} \to Z^{*}\to ZH$) process
 which associated with the $Z$ boson and the $Z$ mainly decays to $q\bar{q}$ pair,
 as shown in Fig.~\ref{fig:ZH_diagram},
 around the $ZH$ production threshold energy shown in Fig.~\ref{fig:ZH_xsec} (a).

 Since Higgs boson mainly decays to $b\bar{b}$ pair at the Higgs mass below 140~GeV region
 as shown in Fig.~\ref{fig:ZH_xsec} (b),
 the final state of the $ZH \to q\bar{q}H$ process forms the four-jet.
 In ILC experiment, the most of interesting physics processes including $ZH$ process
 form the multi-jets final state from the decay of gage bosons ($W, Z$) and heavy flavor quarks ($b, c$),
 thus ILC detectors are required to have the good jet energy resolution for the precise measurement.
 There are three detector concepts, SiD, ILD and $4^{th}$ for the ILC detector, 
 and ILD is the merged concepts of the previous GLD~\cite{GLDDOD} (Asian group)
 and LDC~\cite{LDC} (European group) models for the Letter of Intent (LOI) submission~\cite{ILDLOI}.
 In order to achieve the best jet energy resolution,
 ILD adopt the Particle Flow Algorithm (PFA) suited detector design.
 Since the PFA performance is degraded by the cluster overlapping
 and the double-counting of the particles energy in the calorimeter,
 particles separation in the calorimeter is an important key for better PFA performance.

 The figure-of-merit of the PFA performance from each detector parameter
 relating to the particles separation in the calorimeter is described as
 $F.O.M. = BR^{2}/\sqrt{\sigma^{2}+{R_{M}}^{2}}$,
 where $B$ is a magnetic field,
 $R$ is a detector radius, $\sigma$ is a segmentation of the calorimeter
 and $R_{M}$ is a effective Moliere radius of the calorimeter.
 In order to maximize the $F.O.M.$,
 ILD detector adopts the large radius tracker and high granularity calorimeter
 with $3.5$~T magnetic field.
 In this analysis,
 we study the direct reconstruction of the Higgs boson mass
 with the full detector simulation for $ZH \to q\bar{q}H$, $H\to b\bar{b}$ four-jet mode
 with the ILD detector model.

 \section{Simulation tools}

\begin{figure}[htbp]
 \begin{center}
 \begin{minipage}{0.4\textwidth}
  \begin{center}
   \includegraphics[width=\textwidth]{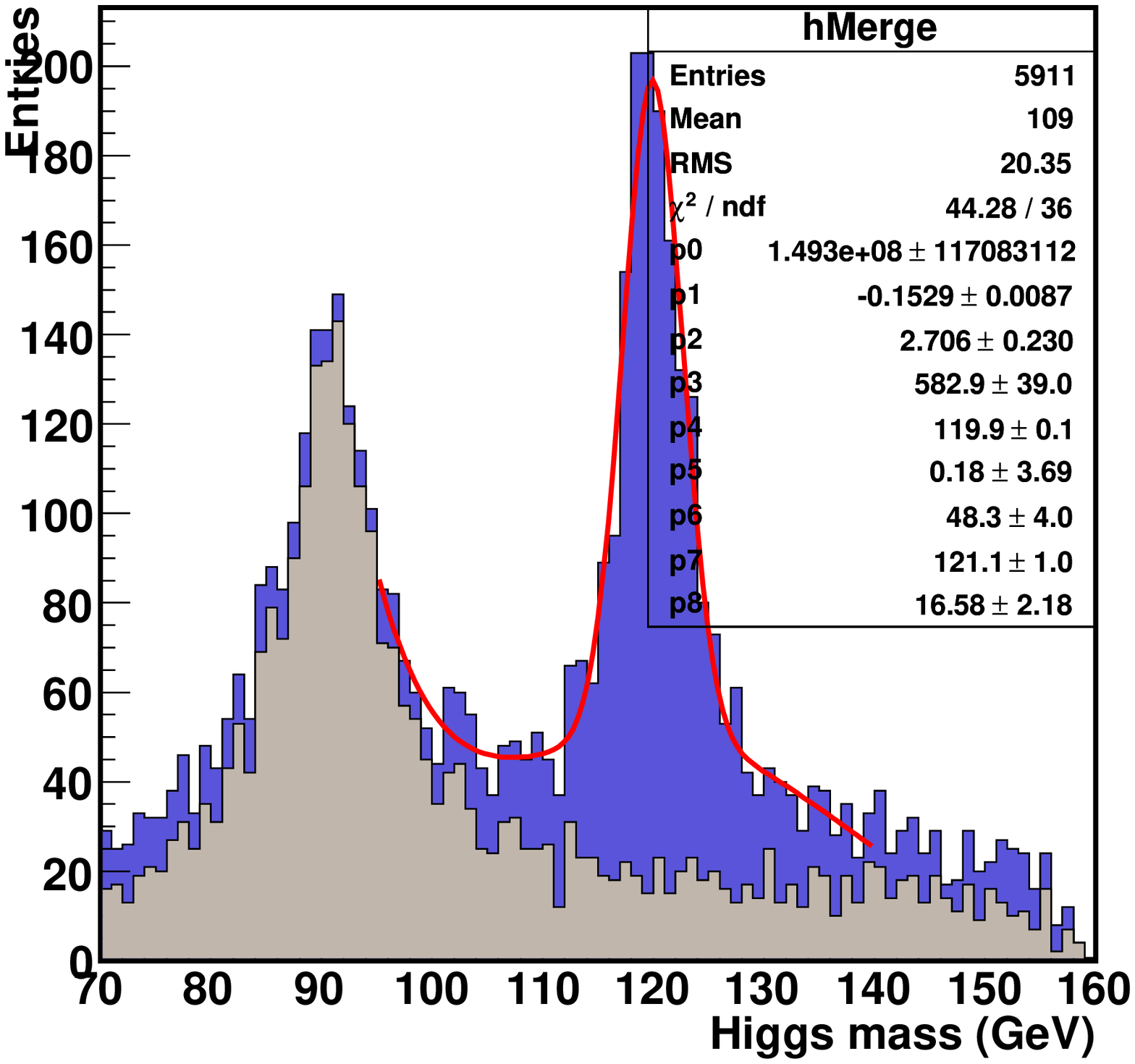}\\
   (a)
  \end{center}
 \end{minipage}
 \begin{minipage}{0.4\textwidth}
  \begin{center}
   \includegraphics[width=\textwidth]{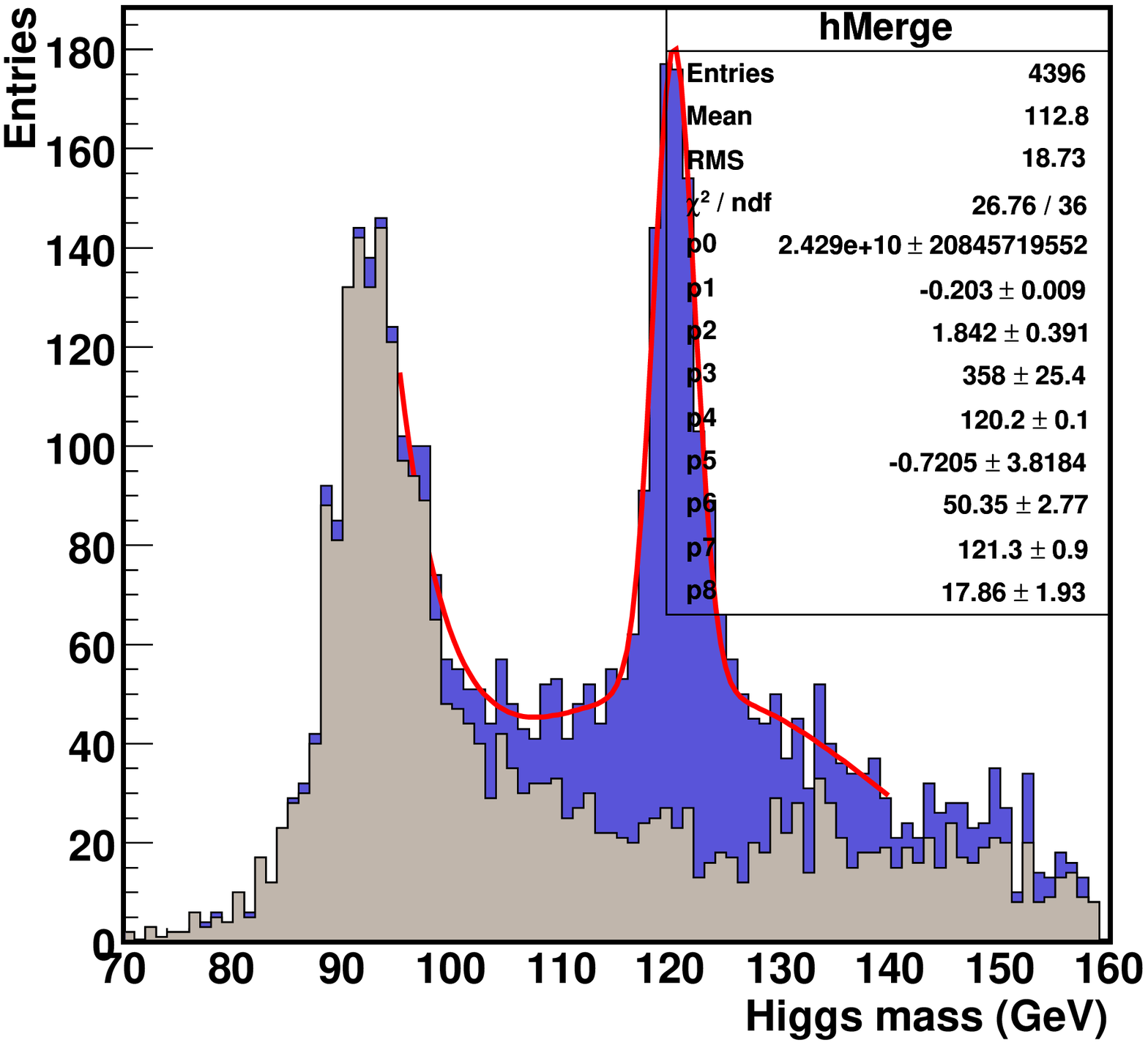}\\   
   (b)
  \end{center}
 \end{minipage}
  \caption{Comparison of the reconstructed Higgs mass distribution for $ZH \to q\bar{q}b\bar{b}$
  only with $ZZ$ background in GLD detector model with the different PFA clustering of
  (a) realistic PandoraPFA and (b) perfect clustering PFA.}
 \end{center}  
  \label{fig:zh_GLD}
 \end{figure}

 For full detector simulation study,
 we use the ILD detector model based Monte Carlo (MC) full simulation package called Mokka,
 which is based on the MC simulation package Geant4~\cite{Geant4}.
 Generated MC hits are reconstructed and smeared in the reconstruction package called MarineReco
 which includes the PFA package called PandoraPFA~\cite{ilcsoft}.
 Since $\sqrt{s}=250~{\rm GeV}$ reconstructed and skimmed signal and background samples
 called DST files are generated for the LOI physics analysis in ILD group,
 we use these DST data samples saved in the linear collider common data format called LCIO.
 For the DST data sample analysis,
 we use the useful analysis package library called Anlib for the event shape analysis
 and jets reconstruction,
 and analysis process is handled through the Root~\cite{Root} based analysis framework
 called JSF~\cite{SimTools}.
 For the comparison of the PFA performance between realistic PFA and perfect-clustering PFA,
 we also use the GLD detector model MC full simulator called Jupiter~\cite{Jupiter}
 with the generating the signal and background events by PYTHIA,
 and reconstruction package called Satellites~\cite{Satellites} based on Root,
 both of them are also controlled in the JSF framework.
 From the comparison of the $ZH \to q\bar{q}H$ in GLD detector model,
 shown in Fig~\ref{fig:zh_GLD},
 PandoraPFA reconstruction performance (a) achieve the comparable performance with perfect-clustering PFA (b)
 in terms of the reconstructed Higgs mass distribution width of $\sigma$
 which corresponds to the jet energy resolution even only the $ZZ \to q\bar{q}q'\bar{q'}$ background is considered.
 Therefore, we shift to the full SM background analysis with common DST data.
 
 \section{Analysis Procedure of $ZH \to q\bar{q}H$ mode} 

 \subsection{MC samples}

 \begin{wrapfigure}{r}{0.38\textwidth}
  \begin{center}
   \includegraphics[width=0.3\textwidth]{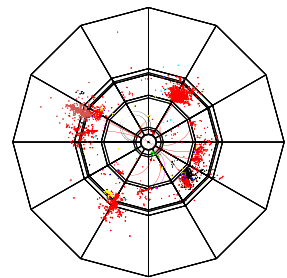}
  \end{center}
  \caption{Typical event display of the $ZH \to q\bar{q}H$ four-jet final state.}
  \label{fig:Evt_disp}  
 \end{wrapfigure}

 The SM Higgs boson is mainly produced through the Higgs-strahlung $e^{+}e^{-} \to ZH$ process
 around the production threshold center-of-mass energy $(\sqrt{s} \sim 230~{\rm GeV})$.
 Since the main decay mode at $M_{H} < 2 M_{W}$,
 Higgs boson mainly decays to $b\bar{b}$ pair,
 thus largest production cross-section is obtained from the $ZH \to q\bar{q}b\bar{b}$ process,
 which forms four-jet final state and both $Z$ and $H$ can be reconstructed directly.
 Fig.~\ref{fig:Evt_disp} shows the typical event display of the $ZH \to q\bar{q}H$ in JSF.
 In this analysis, we assume the center-of-mass energy as the $ZH$ production threshold of
 $\sqrt{s}=250~{\rm GeV}$ and the light Higgs mass of $M_{H} = 120~{\rm GeV}$.
 Each DST data samples is scaled to the integrated luminosity of $\mathcal{L} =250~{\rm fb^{-1}}$
 and the beam polarization to $P(e^{+}, e^{-})=(30\%, -80\%)$.
 The main backgrounds for $ZH \to q\bar{q}b\bar{b}$ are considered as following processes:
 $ZH \to Z^{*}/\gamma \to q\bar{q}$, $e^{+}e^{-} \to WW/ZZ \to qq'q''q'''~{\rm or}~q\bar{q}q'\bar{q'}$, $e^{+}e^{-} \to WW \to {\nu}{\ell}qq'$
 and $e^{+}e^{-} \to ZZ \to {\ell}{\ell}{\ell}{\ell}$.
 Generated signal and background MC samples which scaled to be $\mathcal{L}=250~{\rm fb^{-1}}$
 are summarized in Table.~\ref{table:MC_samples}.
 
 \begin{table}[htpb]
  \begin{center}
   \begin{tabular}{|c|c|c|c|c|c|}
    \hline
    MC samples ($\mathcal{L}=250~{\rm fb^{-1}}$)& $ZH \to qqH$ (sig) &  $qqqq$ & ${\nu}{\ell}qq$ & $\ell\ell\ell\ell$ & $qq$\\
    \hline\hline
    Number of generated events & 51763 & 814163 & 302807 & 98127 & 2529928\\
    \hline
   \end{tabular}
  \end{center}
  \caption{Generated signal and background MC data samples scaled with $\mathcal{L}=250~{\rm fb^{-1}}$.}
  \label{table:MC_samples}
 \end{table}
 In order to correct the escape energy from the heavy quark decay including neutrinos,
 kinematic five constraint (5C) fit is applied,
 which consists of the four constraints (4C) of momentum balance ($\sum{P_{x,y,z}}_i=0$)
 and jets energy balance ($\sum E_{i}-\sqrt{s}=0$) of the four-jet
 and one $Z$ mass constraints for $Z$ candidate di-jet.
 For the kinematic fitting,
 jet energies ($E_{j}$) and jet angles ($\theta$, $\phi$) of each jet are used as measured variables.
 Finally, reconstructed Higgs mass distribution is fitted with the Gaussian convoluted with Gaussian function
 for the signal and exponential function for the contribution from background events
 which remain after the Higgs boson selections.
 
 \subsection{Jet Reconstruction}

 Since the final state of the $ZH \to q\bar{q}H$ mode forms four-jet,
 after the PandoraPFA clustering,
 forced four-jet clustering based on Durham jet-clustering algorithm has applied.
 In order to select the best jet pair combination from the four-jet,
 following $\chi^2$ value is evaluated,
 \begin{equation} \label{chi2:eq}
  \chi^{2}=
   \left(
    \frac{M_{12}-M_{Z}}{\sigma_{M_{Z}}}
   \right)^{2} +
   \left(
    \frac{MissM_{34}-M_{Z}}{\sigma_{MM_{H}}}
   \right)^{2}
 \end{equation}
 where $M_{12}$ is $Z$ candidate di-jet mass,
 $MissM_{34}$ is a missing mass of the remaining Higgs candidate di-jet,
 $M_{Z}$ is the $Z$ boson mass (91.2~GeV),
 and $\sigma_{M_{Z}}$ and $\sigma_{MissM_{34}}$ are
 sigma of distribution of the reconstructed $Z$ boson mass
 and the missing mass of the Higgs candidate jets, respectively.
 In order to select the best jets pair combination,
 $\chi^{2} < 10$ is required for the reconstructed jets pair.

 \subsection{Event selection}

 After the $\chi^{2}$ cut to select the best jet pair combination,
 following event selections are applied for background rejection:
 \begin{enumerate}
  \renewcommand{\labelenumi}{(\alph{enumi})}  
  \setlength{\itemsep}{0pt}
  \item{visible energy : $200 \leq E_{vis} \leq 270~{\rm GeV}$};
  \item{Longitudinal momentum of the $Z$ : $|{P_{\ell}}_{Z}| < 70~{\rm GeV}$ to reduce $ZZ$ background};
  \item{Higgs production angle : $|\cos{\theta_{H}}|<0.85$} to reduce the $ZZ$ background;
  \item{thrust angle : $thrust < 0.9$};
  \item{Number of particles: $N_{particle}>40$ to suppress the ${\ell}{\ell}{\ell}{\ell}$ background};
  \item{Maximum and minimum jet energy fraction: $E_{min}/E_{max}>0.25$};
  \item{Maximum momentum of jet: ${P_{j}}_{max}<100~{\rm GeV}$};
  \item{Y Plus  : $Y Plus  > 0.0001$};
  \item{Y Minus : $Y Minus > 0.001$};
  \item{Minimum angle of $Z$-$H$ jets : $ 20 < {\theta_{ZHj}}_{min} < 135$};
  \item{Maximum angle of $Z$-$H$ jets : $110 < {\theta_{ZHj}}_{max}$};
  \item{$b$-tagging : $P_{btag}>0.5$ from LCFIVTX package}.
 \end{enumerate}

\begin{figure}[htbp]
  \begin{minipage}{0.5\textwidth}
   \includegraphics[width=\textwidth]{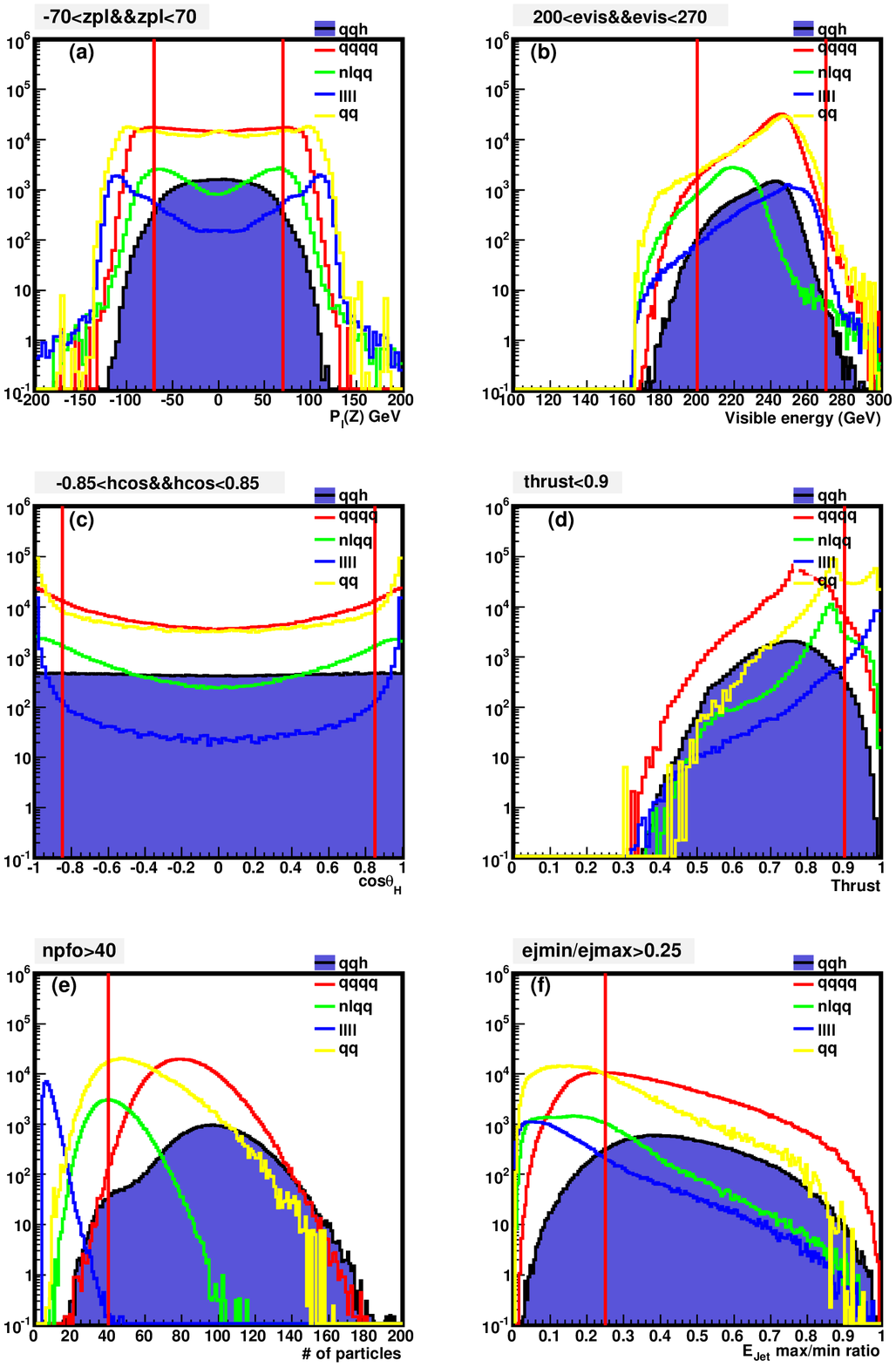}
  \end{minipage}
  \begin{minipage}{0.5\textwidth}
   \includegraphics[width=\textwidth]{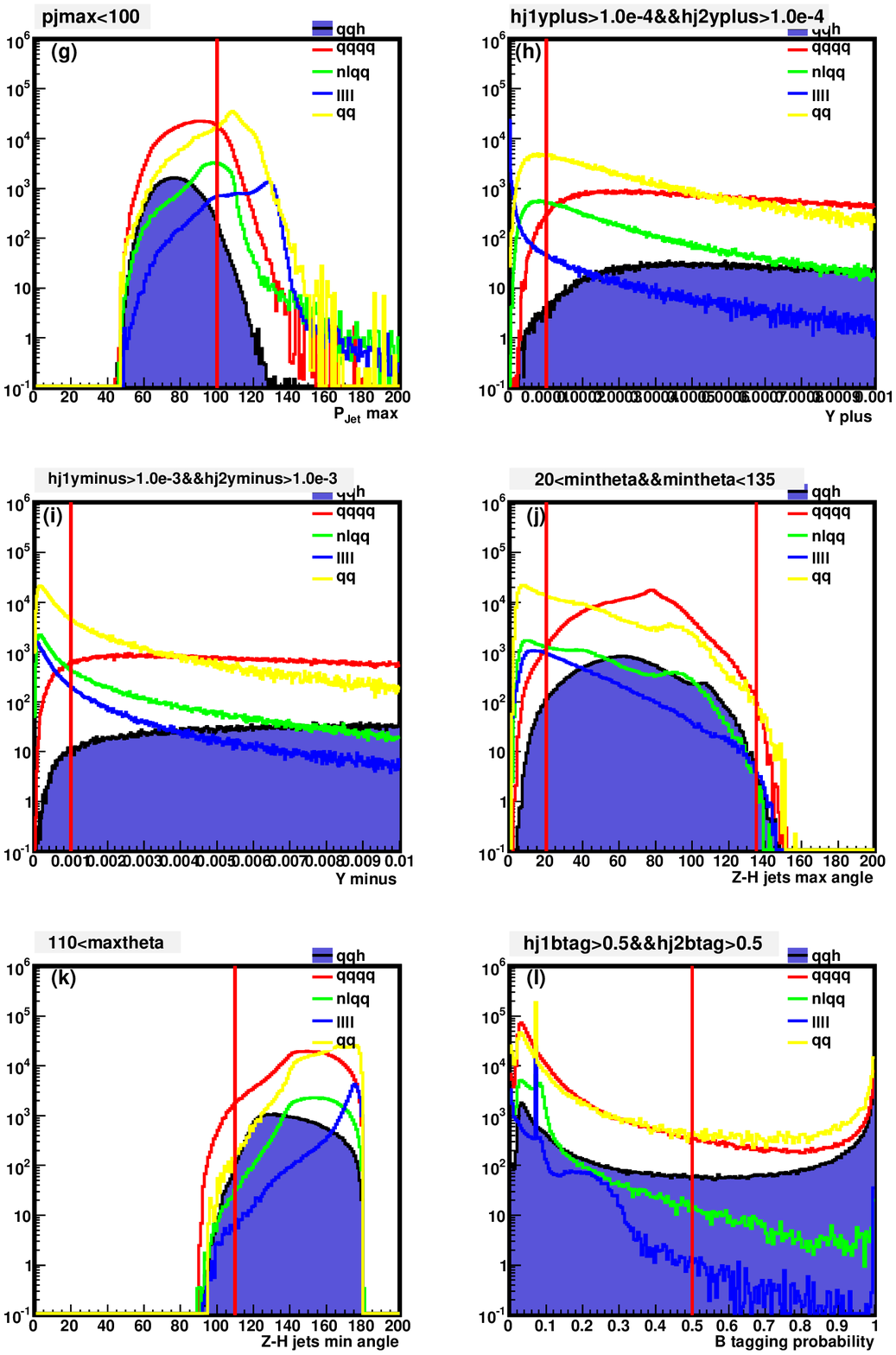}
  \end{minipage}
  \caption{Distribution of each selection variable and its cut positions to select $ZH \to q\bar{q}b\bar{b}$ event.}
  \label{fig:cuts}
 \end{figure}

 The distribution and its cut positions for each selection variable are shown in Fig.~\ref{fig:cuts}.
 Since the $W/Z$ generated in the $WW/ZZ$ background event are relatively boosted
 compare to the $Z$ generated in $ZH$ signal event,
 longitudinal momentum of $Z$ (${P_{\ell}}_{Z}$) and maximum momentum in jets (${P_{j}}_{max}$)
 are higher in $WW/ZZ$ background event than in signal event.
 None jet-like background events are reduced by the number of particles ($N_{PFO}$) cut.
 Y Plus and Y Minus values are threshold Y-values used in the jet clustering topology
 which reconstructed from four-jet to five-jet or three-jet, respectively.
 Minimum and maximum angles between $Z$ and $H$ candidate jets
 are also used for the separation by the event shape difference between $ZH$ event and backgrounds.
 
 Finally, we apply the vertex tagging selection for the neural net output of the $b$-likeness
 analyzed in the vertexing package called LCFIVTX in ilcsoft.
 The reduction summary in each event selection is listed in the Table~\ref{table:zh2qqh}.

 \begin{table}[htbp]
  \scriptsize{
  \begin{center}
   \begin{tabular}{|l|r|r|r|r|r|}
    \hline
    Selections & $ZH \to q\bar{q} H$(Sig) & $qqqq$ & ${\nu}{\ell}qq$ & ${\ell}{\ell}{\ell}{\ell}$ & $qq$\\
    \hline\hline
    no cuts                       & 51745            & 814162            & 302807 & 98127            & 2529928           \\ \hline
    $\chi^{2}$                    & 36748 (71.02 \%) & 688703 (84.59 \%) & 19043 (6.29 \%) & 25375 (25.86 \%) & 541852 (21.42 \%) \\ \hline
    $|{P_{l}}_{Z}|$               & 34952 (67.55 \%) & 479403 (58.88 \%) & 12832 (4.24 \%) & 5565  (5.67 \%)  & 293883 (11.62 \%) \\ \hline
    $E_{vis}$                     & 34924 (67.49 \%) & 477994 (58.71 \%) & 12457 (4.11 \%) & 5335  (5.44 \%)  & 287324 (11.36 \%) \\ \hline
    $|\cos{\theta_{H}}|$          & 30451 (58.85 \%) & 397270 (48.79 \%) & 9934  (3.28 \%) & 2167  (2.21 \%)  & 223873 (8.85 \%)  \\ \hline
    $thrust$                      & 29916 (57.81 \%) & 389703 (47.87 \%) & 8312  (2.75 \%) & 1422  (1.45 \%)  & 103283 (4.08 \%)  \\ \hline
    $N_{particles}$               & 29820 (57.63 \%) & 389514 (47.84 \%) & 4353  (1.44 \%) & 0     (0.00 \%)  & 87022  (3.44 \%)  \\ \hline
    ${E_{j}}_{min}/{E_{j}}_{max}$ & 27843 (53.81 \%) & 297580 (36.55 \%) & 1603  (0.53 \%) & 0     (0.00 \%)  & 40880  (1.62 \%)  \\ \hline
    ${p_{j}}_{max}$               & 27622 (53.38 \%) & 289490 (35.56 \%) & 1500  (0.50 \%) & 0     (0.00 \%)  & 31382  (1.24 \%)  \\ \hline
    $Y plus$                      & 27607 (53.35 \%) & 288421 (35.43 \%) & 1465  (0.48 \%) & 0     (0.00 \%)  & 30773  (1.22 \%)  \\ \hline
    $Y minus$                     & 27559 (53.26 \%) & 287825 (35.35 \%) & 1354  (0.45 \%) & 0     (0.00 \%)  & 27250  (1.08 \%)  \\ \hline
    ${\theta_{Z-H j}}_{min}$      & 27311 (52.78 \%) & 285704 (35.09 \%) & 1284  (0.42 \%) & 0     (0.00 \%)  & 24601  (0.97 \%)  \\ \hline
    ${\theta_{Z-H j}}_{max}$      & 27031 (52.24 \%) & 277203 (34.05 \%) & 1263  (0.42 \%) & 0     (0.00 \%)  & 24280  (0.96 \%)  \\ \hline\hline
    $b-tagging$                   & 5972  (11.54 \%) & 4732   (0.58  \%) & 0     (0.00 \%) & 0     (0.00 \%)  & 458    (0.02 \%)  \\
    \hline
   \end{tabular}
  \end{center}
  }
  \caption{Backgrounds reduction summary in each selection for $ZH \to q\bar{q}b\bar{b}$.}
  \label{table:zh2qqh}
 \end{table}

 From the reduction summary of Table.~\ref{table:zh2qqh},
 ${\ell}{\ell}{\ell}{\ell}$ four-leptonic background can be suppressed completely
 by number of particles cut ($N_{PFOs}<40$) and the remaining backgrounds
 are $qqqq$ and $qq$ which including $b$-quarks event after applying the $b$-tagging.

 \section{Results}

 \begin{wrapfigure}{r}{0.40\textwidth}
  \includegraphics[width=0.40\textwidth]{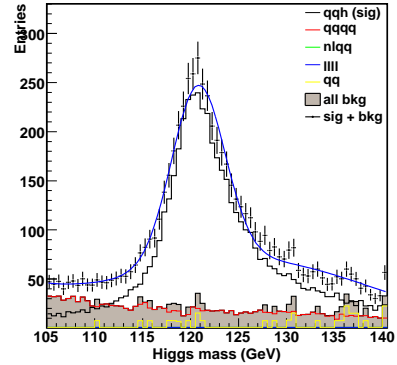}
  \caption{Reconstructed Higgs mass distribution of $ZH \to q\bar{q}b\bar{b}$.}
  \label{fig:zh2qqbb}
 \end{wrapfigure}
 Reconstructed Higgs mass distribution after the selection of $ZH \to q\bar{q}b\bar{b}$
 is fitted with the function of Gaussian convoluted Gaussian with the exponential function assuming the background,
 as shown in Fig.~\ref{fig:zh2qqbb}.
 Fitted results of the reconstructed $ZH \to q\bar{q}b\bar{b}$ Higgs mass distribution
 are summarized in the Table.~\ref{talble:results}.
 From the fitted results,
 Higgs mass (${M_{H}}=120~{\rm GeV}$ at MC) is reconstructed as $M_{H} = 120.79 \pm 0.089 {\rm GeV}$
 and the measurement accuracy of cross-section to $ZH \to q\bar{q}b\bar{b}$
 is obtained as $\delta\sigma/\sigma=5.3\%$.

 \begin{table}[b]
  \begin{center}
  \begin{tabular}{|c|c|}
   \hline
   Higgs mass (${M_{H}}=120~{\rm GeV}$ at MC) & $M_{H} = 120.79~(\rm GeV)$\\
   \hline
   Measurement accuracy of $M_{H}$ & $\delta M_{H}=89~(\rm MeV)$\\
   \hline
   Measurement accuracy of $\sigma(ZH\to q\bar{q}b\bar{b})$ & $\delta\sigma/\sigma=5.3\%$\\
   \hline
  \end{tabular}
  \end{center}
  \caption{Fitted results for the reconstructed Higgs mass distribution.}
  \label{talble:results}
 \end{table}
 
 \section{Conclusion}

 Simulation study of the direct reconstruction of the Higgs boson in
 $ZH \to q\bar{q}b\bar{b}$ four-jet mode with the Higgs mass of 120~GeV
 at the $\rm \sqrt{s}=250~GeV$ and the integrated luminosity of $\mathcal{L}=250 fb^{-1}$ has performed
 for the ILD detector model considering with the $qqqq,~{\nu}{\ell}qq~{\ell}{\ell}{\ell}{\ell},~qq$ background processes.
 From the study, measurement accuracy of the reconstructed Higgs mass is estimated as $87~{\rm MeV}$
 and the measurement accuracy of the cross-section of $ZH \to q\bar{q}b\bar{b}$ mode
 is obtained as $\delta\sigma/\sigma = 5.3\%$.
 
 \section*{Acknowledgment}

 I would like to thank to everyone who join the ILC physics WG subgroup~\cite{physwg}
 for useful discussion of this work
 and to ILD optimization group members who maintain the softwares and MC samples.
 This study is supported in part by the Creative Scientific Research Grant No.~18GS0202
 of the Japan Society for Promotion of Science and promotion.



{\footnotesize
 \begin{thebibliography}{99}
  \bibitem{RDR} ILC Reference Design Report (RDR) http://www.linearcollider.org/rdr/
  \bibitem{LEP} The LEP Electroweak Working Group, arXiv:0811.4682 [hep-ex] (November 2008).
  \bibitem{CDF} CDF Collaboration and D0 Collaboration,
	  arXiv:0903.4001 [hep-ex].
  \bibitem{GLDDOD}   GLD Detector Outline Document (DOD), arXiv:physics/0607154v1 [physics.ins-det]
  \bibitem{LDC}      http://ilcldc.org/documents/dod
  \bibitem{ILDLOI}   http://www.ilcild.org/
  \bibitem{Geant4}   GEANT4 Collaboration: S Agostinelli et al, Nucl. Instrum. Methods A506, 250 (2003).
  \bibitem{ilcsoft}  http://ilcsoft.desy.de/portal/
  \bibitem{Root}     http://root.cern.ch/
  \bibitem{SimTools} http://acfahep.kek.jp/subg/sim/simtools/
  \bibitem{Jupiter}  ACFA Linear Collider Working Group, KEK Report 2001-11, August, 2001.
  \bibitem{Satellites}  Proceedings of the APPI Winter Institute, KEK Proceedings 2002-08, July (2002).
  \bibitem{cheated_PFA} S.~Yamamoto, K.~Fujii and A.~Miyamoto, arXiv:0809.4111 [physics.comp-ph].
  \bibitem{physwg}   http://www-jlc.kek.jp/subg/physics/ilcphys/
 \end{thebibliography}
}
\end{document}